\documentclass[%
 reprint,
 amsmath,amssymb,
 aps,
prb,
floatfix,
]{revtex4-2}

\usepackage{graphicx} 

\usepackage{float}
\usepackage{placeins}
\usepackage{xcolor}
\usepackage{lipsum}
\usepackage{amsmath}
\usepackage{physics}
\usepackage{dcolumn}% Align table columns on decimal point
\usepackage{bm}% bold math
\usepackage{hyperref}
\usepackage{times}

\begin{document}

\preprint{APS/123-QED}

\title{
Non-Ohmic behavior in (Bi$_{1-x}$Sb$_x$)$_2$Te$_3$ by Joule heating
}
\author{Sofie K\"olling\textsuperscript{1}}
\author{Daan H. Wielens\textsuperscript{1}}%
\author{\.{I}nan\c{c} Adagideli\textsuperscript{1, 2}}
\author{Alexander Brinkman\textsuperscript{1}}
\affiliation{\textsuperscript{1}MESA+ Institute for Nanotechnology, University of Twente, Enschede, The Netherlands}
\affiliation{\textsuperscript{2}Faculty of Engineering and Natural Sciences, Sabanci University, Turkey}
\date{\today}
\newcommand{\supplementarysection}{%
  \setcounter{figure}{0}% Reset figure counter
  \let\oldthefigure\thefigure% Capture figure numbering scheme
  \renewcommand{\thefigure}{S\oldthefigure}% Prefix figure number with S
  \section*{Supplementary information}% Set supplementary section
  }

\begin{abstract}
A prerequisite to using the net spin polarization generated by a source-drain bias in three-dimensional topological insulators for spintronic applications, is understanding how such a bias alters the transport properties of these materials.
At low temperatures, quantum corrections can dominate the temperature dependence of the resistance. Although a DC bias does not break time-reversal symmetry and is therefore not expected to suppress quantum corrections, an increase of the electron temperature due to Joule heating can cause a suppression. This suppression at finite bias can lead to a non-Ohmic differential resistance in the three-dimensional topological insulator (Bi$_{1-x}$Sb$_x$)$_2$Te$_3$, consisting of a zero-bias resistance peak (from electron-electron interactions) and a high-bias background (from weak antilocalization). We show that the bias voltage dependence of quantum corrections can be mapped to the temperature dependence, while the heating effect on the lattice temperature remains small. When searching for non-Ohmic effects due to novel phenomena in three-dimensional topological insulators, Joule heating should not be overlooked.
\end{abstract}
\maketitle

\section{Introduction}
Three-dimensional topological insulators host conducting surface states, while the bulk remains insulating. These surface states are spin-momentum locked. This generates a nonzero spin polarization upon applying a DC source-drain bias, without requiring an external magnetic field. Possible applications of this mechanism are magnetic switching for memory operations \cite{liu2012spin, mellnik2014spin, khang2018conductive} or energy storage via nuclear spins \cite{bozkurt2018Work, bozkurt2023Topological, bozkurt2024entropy}.

Such applications require a thorough understanding of electronic transport in topological insulators under a DC bias. At low temperatures, the conductivity of a topological surface state is governed by quantum corrections: weak antilocalization (WAL) increases the conductance, and electron-electron interactions (EEI) decrease the conductance. Generally, EEI dominate over WAL \cite{jing2016weak, lu2014finite, kuntsevich2016low}, causing the conductivity to decrease when lowering the temperature. In 2D systems such as topological surface states, both WAL and EEI scale logarithmically with temperature \cite{gantmakher2005electrons}. However, contrarily to WAL, EEI are robust in magnetic field \cite{lee1985disordered}.

Theoretically, a DC bias should not influence the quantum corrections to conductance, because time-reversal symmetry is maintained \cite{lee1985disordered}. In contrast, experiments show measurable effects of a DC source-drain bias on quantum corrections: for instance through dispersion decoherence \cite{somphonsane2020universal}, an Overhauser field originating from nuclear polarization \cite{jiang2020dynamic}, or Joule heating \cite{gusev2019electronic, nandi2018logarithmic}. The latter effect is expected to occur in any resistive system and therefore cannot be omitted. 

The theory of Joule heating in disordered metals has been described by Abrahams and Anderson \cite{anderson1979possible, abrahams1980non}. At low temperatures, electron-phonon interactions become less effective. Therefore, thermal contact between electrons and the lattice is reduced.
Upon applying an electric field $E$, electrons heat up in-between inelastic scattering events (on a timescale of $\tau_\phi$) before they are able to transfer the absorbed heat to the lattice, and thereby the electron temperature is raised. 
The heating power equals $\sigma E^2$, where $\sigma$ is the conductivity.
If the applied electric field is sufficiently large, it will dominate over the influence of lattice temperature in transport experiments. 

A governing energy $\epsilon$ can be defined, which is either set by temperature or voltage as $\epsilon \in \mathrm{max}(T, V^{2/(2+p)})$ \cite{anderson1979possible}. 
Here, $p$ is a parameter setting the dephasing $\tau_\phi \sim T^{-p}$. The EEI and WAL conductivity corrections scale logarithmically with $\epsilon$, so the newly introduced dependence on voltage (via $\epsilon$) implies non-Ohmic signatures when quantum corrections are suppressed by Joule heating.

In this study, we investigate the non-Ohmic differential resistance in the three-dimensional topological insulator (Bi$_{1-x}$Sb$_x$)$_2$Te$_3$ (BST). 
Zero-bias resistance peaks have been attributed to EEI in these materials, however, the mechanism which suppresses EEI at finite bias remains unclear \cite{stehno2017conduction, janssen2024characterization}.
Our experiments are similar to previous work on logarithmic conductivity corrections due to Joule heating in topological insulators \cite{nandi2018logarithmic}. We extend this picture by finding that the differential resistance contains two non-Ohmic contributions. Firstly, we measure a zero-bias resistance peak, which is robust in an external magnetic field. Secondly, the differential resistance increases at high bias, and this contribution is suppressed in a perpendicular magnetic field.  We investigate whether both non-Ohmic contributions can be related to the suppression of EEI and WAL by Joule heating.

\section{Materials and methods}
We study the transport properties of BST by fabricating Hall bar devices {(see \cite{geometry_app} for a schematic depiction of the device geometry and measurement setup)}. This allows us to determine the longitudinal ($R_{xx}$) and Hall ($R_{xy}$) resistance {by sourcing a DC source-drain bias ($I_\mathrm{DC}$) and measuring the voltages ($V_{xx}$, $V_{xy}$)}.
BST thin films with an antimony content of $x \approx 0.6$ are deposited on vicinal Al$_2$O$_3$ substrates using molecular beam epitaxy (MBE) \cite{mulder2023enhancement}. The film thickness is approximately 10 nm. Next, the films are contacted with Ti/Pd contacts {(thicknesses 5 nm/35 nm)} and are Ar-milled into Hall bar devices of varying length ($L$, between voltage probes) and width ($W$, across the channel). Finally, the devices are capped with an AlO$_x$ layer (thickness 30 nm) by means of low-temperature atomic layer deposition (ALD) at $T = 100$ $^{\circ}$C. An overview of the measured devices is provided in Table~\ref{tab:devices}.

Transport measurements are performed at cryogenic temperatures in a physical properties measurement system (PPMS) using magnetic fields up to 9 T. A DC source-drain current is applied through the current leads using a Keithley 2400 SourceMeter, and the corresponding voltages are measured in a 4-terminal configuration by a Keitley 2182A Nanovoltmeter. Because Joule heating is governed by the electric field, we will analyze the data in terms of the measured bias voltage instead of the applied bias current.
\begin{figure*}
    \centering
    \includegraphics[width = 0.98\textwidth]{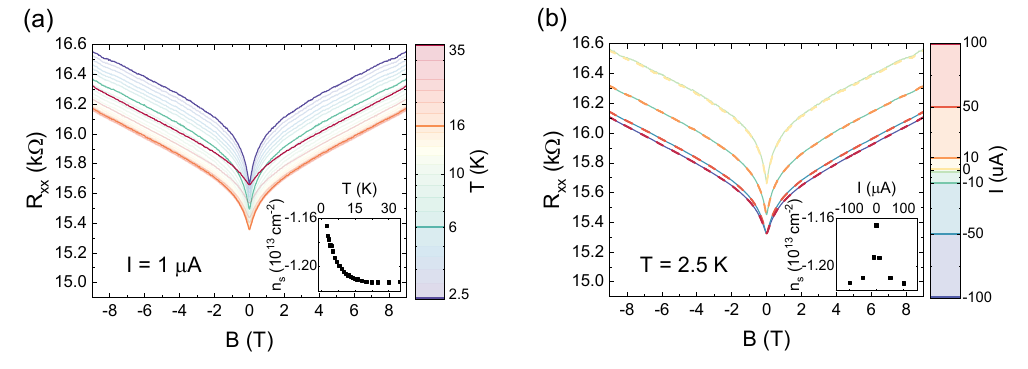}
    \caption{Longitudinal resistance in a (Bi$_{1-x}$Sb$_x$)$_2$Te$_3$ Hall bar (D1) as function of (a) temperature at $I_\mathrm{DC} = 1$ $\mu$A and (b) DC bias current at $T = 2.5$ K. The traces obtained using $I_\mathrm{DC}>0$ and $I_\mathrm{DC}<0$ in (b) overlap, results for $I_\mathrm{DC}>0$ are plotted using dashed lines for clarity. The graph insets show the extracted {charge-carrier densities} using a single-band Drude model.}
    \label{fig:MR_IvsT}
\end{figure*}

\begin{table}%The best place to locate the table environment is directly after its first reference in text
\caption{\label{tab:devices}%
Device overview.}
\begin{ruledtabular}
\begin{tabular}{l c c c r r}
\textrm{Sample}&
\textrm{Material}&
\textrm{W ($\mu$m)}&
\textrm{L ($\mu$m)}&
\textrm{$x$}&
\textrm{$y$}
\\
\colrule
D1     & (Bi$_{1-x}$Sb$_x$)$_2$Te$_3$          & 6          & 60  & 0.6 & -     \\
D2     & (Bi$_{1-x}$Sb$_x$)$_2$Te$_3$          & 5          & 50  & 0.6 & -     \\
D3     & V$_y$(Bi$_{1-x}$Sb$_x$)$_{2-y}$Te$_3$ & 20         & 400 & 0.7 & 0.01     \\
D4     & (Bi$_{1-x}$Sb$_x$)$_2$Te$_3$          & 40         & 400 & 0.6 & -       
\end{tabular}
\end{ruledtabular}
\end{table}

\section{Magnetotransport}
In the following sections, we analyze dephasing effects upon applying a nonzero bias current. The aim is to find whether non-Ohmic signatures can be directly attributed to Joule heating.
To obtain insight into the dephasing mechanisms in the Hall bar devices, the magnetoresistance in an out-of-plane external magnetic field is studied. First, in Fig.~\ref{fig:MR_IvsT}(a) the magnetoresistance is measured at varying temperatures using $I_\mathrm{DC} = 1$ $\mu$A, to exclude high-bias effects. 

The traces contain a characteristic weak antilocalization cusp that is suppressed at high temperature, on top of a linear background typical for topological insulators \cite{deboer2019magnetoresistance}. Next, in Fig.~\ref{fig:MR_IvsT}(b) the magnetoresistance is measured at varying DC bias {current} and $T = 2.5$ K, to exclude high-temperature effects. 
{For low currents, the entire magnetoresistance trace lies at a higher resistance value than the traces obtained at higher currents. The traces obtained at positive currents are dashed, to show their overlap with the negative counterpart.}

The two datasets resemble each other: upon increasing temperature or voltage, the average resistance is suppressed, and the weak antilocalization cusps broaden. A similar correspondence between elevated temperature and bias is reflected in the {charge-carrier densities} extracted from $R_{xy}$ using a single-band Drude model (Fig.~\ref{fig:MR_IvsT}(a) and (b) insets). Further magnetoresistance data is provided in \cite{mr_app}\nocite{altshuler1980magnetoresistance}. 

It is tempting to conclude that applying a bias current straightforwardly increases the temperature in our system. 
However, upon closer investigation, slight differences occur. For instance, the resistance at $B = 0$ T in Fig.~\ref{fig:MR_IvsT}(a) {is a non-monotonic function of temperature with a minimum} around 15 K, but this effect {is less pronounced in the $B = 0$ resistance} as a function of bias in Fig.~\ref{fig:MR_IvsT}(b), even though the {charge-carrier densities} attain similar values at high temperature/bias. {The differences become more pronounced upon plotting the differential conductivity ($(W/L)\cdot \dd I/\dd V_{xx}$) for a range of temperatures in Fig.~\ref{fig:sigma_ZBRP_1}, where for instance we see that the conductivity values at the local maxima at $T = 3$ K cannot be accessed if the base temperature is increased.} Consequently, we study the differential resistance as function of bias in more detail and confirm {to what extent} the data follows $\epsilon \in \mathrm{max}(T, V^{2/(2+p)})$.

\section{Zero-bias resistance peak}
\begin{figure}
    \centering
    \includegraphics[width=0.99\columnwidth]{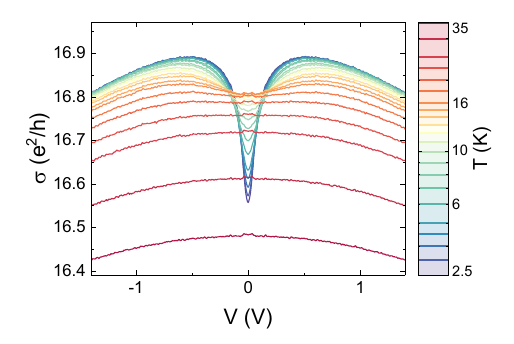}
    \caption{Differential longitudinal conductivity $\sigma = (W/L)\cdot \dd I/\dd V_{xx}$ in device D1 for a varying temperature, plotted as function of the measured $V_{xx}$. The data has been smoothed prior to numerically differentiating.}
    \label{fig:sigma_ZBRP_1}
\end{figure}

The differential resistance is plotted for a range of temperatures in Fig.~\ref{fig:sigma_ZBRP_1} {as a function of bias voltage $V$. Here, and in the subsequent analysis, $V$ corresponds to the measured $V_{xx}$ when sourcing $I_\mathrm{DC}$ through the Hall bar.} Two features are visible: a zero-bias resistance peak (ZBRP), and a gradual increase of the differential resistance at high bias.

The raw data underlying Fig.~\ref{fig:sigma_ZBRP_1} requires filtering to make the non-Ohmic contributions visible. Before numerically differentiating, a Savitsky-Golay filter is applied to the obtained voltage, which we measure while applying a DC current. Although this enhances the visibility of non-Ohmic features, it also reduces the magnitude of the resistance at low bias, especially at lower temperatures where the width of the ZBRP becomes comparable to the Savitsky-Golay window size. To verify the magnitude of the differential conductance at low bias, we compare the filtered and unfiltered data to an $R(T)$ curve measured at $I_\mathrm{DC}=1$ $\mu$A \cite{filt_app}, and conclude that at low temperatures the unfiltered data better reflects the magnitude of the zero bias resistance than the filtered data, but the filtered data enhances the high-bias features.

{Figure~S6(a)} shows the dataset of Fig.~\ref{fig:sigma_ZBRP_1} on a logarithmic scale \cite{zbrp_app}.  
\begin{figure*}[t]
    \centering
    \includegraphics[width=0.95\textwidth]{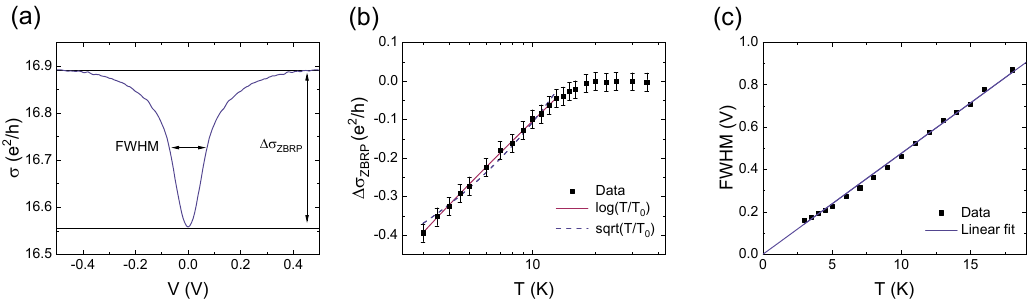}
    \caption{{(a) Differential conductivity obtained at $T = 3$ K, also shown in Fig.~\ref{fig:sigma_ZBRP_1}, in which the full width at half maximum (FWHM) and conductance correction from the zero-bias resistance peak ($\Delta\sigma_\mathrm{ZBRP}$) are indicated.} (b) $\Delta\sigma_\mathrm{ZBRP}$ as a function of temperature extracted from Fig.~\ref{fig:sigma_ZBRP_1}, and comparison with different models. Error bars correspond to the noise level in the raw data. (c) Full width at half maximum extracted from Fig.~\ref{fig:sigma_ZBRP_1}, with a linear fit.}
    \label{fig:sigma_ZBRP_2}
\end{figure*}
Here, it becomes evident that the low-voltage differential resistance saturates at a value set by temperature, beyond which the ZBRP decays logarithmically. As underlined by the magnetotransport data in Fig.~\ref{fig:MR_IvsT}(b), the ZBRP is not suppressed in a magnetic field: {when taking a line-cut at each magnetic field value in Fig.~\ref{fig:MR_IvsT}(b), the ZBRP would be visible. This robustness in magnetic field} could point towards electron-electron interactions being the underlying mechanism. Moreover, Fig.~\ref{fig:sigma_ZBRP_1} emphasizes that voltage and temperature are not interchangeable: some resistance values can only be accessed in a specific voltage-temperature combination.

To study the ZBRP in more detail, {we extract the amplitude $\Delta\sigma_\mathrm{ZBRP}$ as shown in Fig.~\ref{fig:sigma_ZBRP_2}(a), and plot the results as a function of temperature in Fig.~\ref{fig:sigma_ZBRP_2}(b)}. The amplitude decreases logarithmically with temperature, corresponding to two-dimensional EEI \cite{lee1985disordered} {which might be attributed to the surface state. A square-root scaling, expected in a three-dimensional (bulk) system, provides a lesser fit to the data}. 
The data saturates at a nonzero value due to the finite noise level in the unfiltered dataset at high temperatures. 
In {Fig.~S6(b)} the differential resistance is normalized to the ZBRP amplitude and plotted as function of $eV/k_\mathrm{B}T$, collapsing the normalized data to a single curve. This universal scaling corresponds to the full-width at half maximum (FWHM) increasing linearly with temperature in {Fig.~\ref{fig:sigma_ZBRP_2}(c)} (note that we only extract the FWHM for temperatures where $\Delta\sigma_\mathrm{ZBRP}$ is nonzero).
In case of Joule heating, with the conductivity corrections scaling logarithmically as a function of $\epsilon \in \mathrm{max}(T, V^{2/(2+p)})$, the linear FWHM would point towards a scaling of $V \sim T^2$ \cite{zbrp_app}.
% (see appendix \ref{sec:fwhm_theory}).

Several origins for a ZBRP are discussed in literature, including the Kondo effect in Bi$_{1.33}$Sb$_{0.67}$Se$_3$ nanowires \cite{cho2016kondo}, interaction effects between spatially separated quantum Hall states \cite{machida2000anomalously}, helical Luttinger liquids \cite{li2015observation, jia2022tuning}. Our films could show similar effects if surface hybridization would induce a quantum spin Hall state \cite{asmar2018topological}. However, due to the high {charge-carrier density} ($\sim 10^{13}$ cm$^{-2}$) and the non-quantized resistance, we do not expect that the chemical potential lies within a surface hybridization gap, and will investigate whether Joule heating in combination with EEI can explain the measured results.

\section{High-bias resistance}
If the electron temperature is indeed altered by applying a bias voltage, not only EEI but also WAL should result in a non-Ohmic contribution to the differential resistance. 
The source-drain bias dependence of weak (anti)localization has been an object of discussion \cite{lee1985disordered}. In general, a DC bias does not break down WAL. However, if the bias voltage enhances the temperature through Joule heating, WAL can be suppressed through the temperature-dependent change in phase coherence length. 
The following section elaborates on this. 

\begin{figure*}
    \centering
    \includegraphics[width=0.95\textwidth]{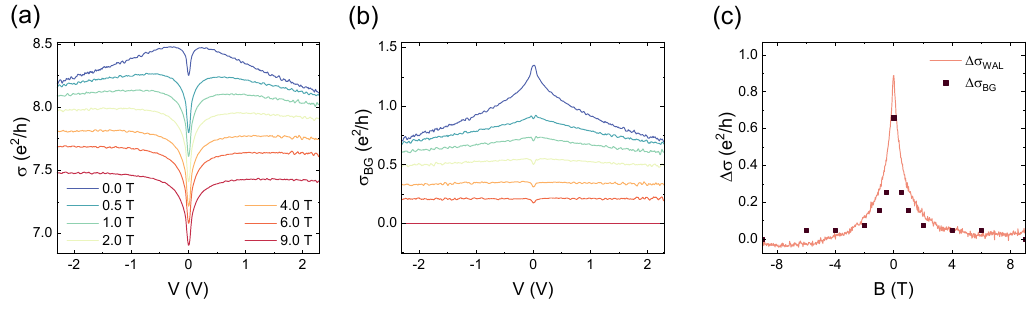}
    \caption{(a) Differential conductivity measured at varying magnetic field strength. The data has been symmetrized in magnetic field, hence only positive field values are shown. (b) Residual background conductivity after writing $\sigma = \sigma_\mathrm{EEI} + \sigma_\mathrm{BG}$ and approximating $\sigma_\mathrm{BG}\approx \sigma - \sigma(9\mathrm{T})$. (c) Total variation in $\sigma_\mathrm{BG}$ extracted from (b) (data points) compared with the WAL conductivity correction extracted from Fig. \ref{fig:MR_IvsT} at $I_\mathrm{DC} = 1$ $\mu$A (solid line), after subtracting the linear background magnetoresistance. Data from device D2, {measured at $T = 2.5$ K}.}
    \label{fig:3_WAL_sigBG_1}
\end{figure*}

Beyond the zero-bias resistance peak, the differential resistance increases gradually with bias voltage. This background contribution is suppressed in an external out-of-plane magnetic field, see Fig.~\ref{fig:3_WAL_sigBG_1}(a).  
Note that this data is obtained on a different sample from the data in Fig.~\ref{fig:MR_IvsT}-\ref{fig:sigma_ZBRP_2}, so although the transport is similar, the data cannot be compared one-to-one. In this dataset, we isolate the background contribution by writing $\sigma = \sigma_\mathrm{EEI} + \sigma_\mathrm{BG}$,
where $\sigma_\mathrm{EEI}$ is the conductance corresponding to the EEI contribution (ZBRP) and $\sigma_\mathrm{BG}$ is the background contribution. Assuming that (i) $\sigma_\mathrm{EEI}$ is not affected by a magnetic field \cite{lee1985disordered} and (ii) $\sigma_\mathrm{BG}$ is fully suppressed at $B = 9$ T, we approximate $\sigma_\mathrm{BG} \approx \sigma - \sigma(9\mathrm{\ T})$. The results are shown in Fig.~\ref{fig:3_WAL_sigBG_1}(b): for magnetic fields exceeding 2 T, $\sigma_\mathrm{BG}$ becomes Ohmic. The remaining $B$-dependent offset corresponds to the non-saturating (linear background) magnetoresistance also visible in Fig.~\ref{fig:MR_IvsT} \cite{deboer2019magnetoresistance}.

Next, we search for an orgin of the non-Ohmic part of $\sigma_\mathrm{BG}$. {The suppression of the non-Ohmic part of $\sigma_\mathrm{BG}$ at high magnetic fields suggests that it should be reflected in the magnetoresistance. Hence, we compare} it with a suppression of weak antilocalization. At each magnetic field value, we compare the total bias-dependent change in $\sigma_\mathrm{BG}$ to the corresponding weak antilocalization correction measured at low bias. 
To this end, we {use the $B = 9$ T value of the magnetoresistance as a reference point for the weak antilocalization correction.}
% assume that weak antilocalization is fully suppressed at $B = 9$ T. 

The total change $\Delta \sigma_\mathrm{BG} = \sigma_\mathrm{BG}(V = 0) - \sigma_\mathrm{BG}(V = V_\mathrm{max})$ is plotted as a function of magnetic field in Fig. \ref{fig:3_WAL_sigBG_1}(c). This way, each $\Delta\sigma_\mathrm{BG}$ data point resembles the background conductivity suppression due to an applied voltage, at fixed magnetic field. In calculating $\Delta \sigma_\mathrm{BG}$ we subtracted the $B$-dependent offset visible in Fig.~\ref{fig:3_WAL_sigBG_1}(b) corresponding to the linear background magnetoresistance. Consequently, we compare $\Delta \sigma_\mathrm{BG}$ to the magnetoresistance obtained at low bias ($I = 0.1\ \mu$A), after subtracting the linear background.
Figure \ref{fig:3_WAL_sigBG_1}(c) shows that the two are similar: hence we attribute the non-Ohmic part of $\sigma_\mathrm{BG}$ to a suppression of weak antilocalization. {Although the resemblance with weak antilocalization is striking, we have not yet confirmed whether the suppression is due to Joule heating, or other mechanisms such as an Overhauser field from nuclear polarization \cite{jiang2020dynamic}.}

\section{Temperature-voltage correspondence}
In the previous sections, we have distinguished two components contributing to the non-Ohmic resistance in BST: a zero-bias resistance peak (EEI) and a non-Ohmic background (WAL). Next, we check whether the full IV curve can be explained by a single effective electron temperature. {If so, both features could be attributed to Joule heating of the electron temperature.}

As we have seen in Fig.~\ref{fig:sigma_ZBRP_1}, not all differential resistances can be accessed by only varying bias voltage or temperature. In a possible scenario, the EEI and WAL terms are altered similarly by a changing electron temperature, but an additional contribution is present which only depends on lattice temperature.  Therefore, we isolate the EEI and WAL contributions from the background by only considering relative changes as function of voltage, and study the temperature-voltage correspondence in each contribution independently.

\begin{figure*}
    \centering
    \includegraphics[width=\textwidth]{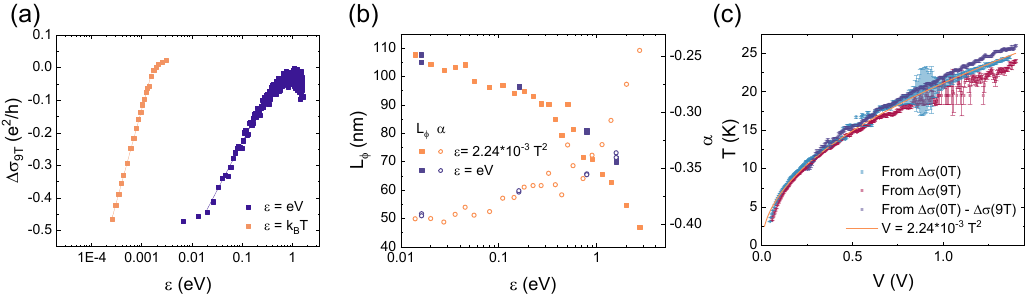}
    \caption{(a) Conductivity correction corresponding to the zero-bias resistance peak ($\Delta \sigma_\mathrm{ZBRP}$) measured at 9 T. $\epsilon$ corresponds to varying voltage (measured at $T = 2.5$ K) or temperature, similar to the 0 T data in Fig. \ref{fig:sigma_ZBRP_2}(b). {Lines originate from linear fits through the data.} (b) $L_\phi$ (solid) and $\alpha$ (open) extracted from fitting Eq. (\ref{eq:hln}) on the datasets of Fig. \ref{fig:MR_IvsT}. The temperature-dependent data is rescaled to $\epsilon = 2.24\cdot 10^{-3}T^2$, from the $T(V)$ curve in (c), to show the correspondence between temperature and voltage. (c) Effective temperature reached upon applying a DC bias at a lattice temperature of $T = 2.5$ K, obtained by comparing temperature- and voltage dependent datasets. At $B = 0$ T, both EEI and WAL contribute, whereas at 9 T only EEI remain. Irrespective of magnetic field, the same $T(V)$ curve holds, showing that WAL and EEI are both equivalently suppressed by Joule heating. Data from device D1.}
    \label{fig:VT_curves}
\end{figure*}

Firstly, to analyze the EEI contribution we track the magnitude of the ZBRP at $B = 9$ T as a function of voltage and temperature in Fig.~\ref{fig:VT_curves}(a). $\Delta \sigma(T)$ corresponds to the ZBRP amplitude, also shown in Fig.~\ref{fig:sigma_ZBRP_2}(b).

Secondly, to analyze the WAL contribution, we also require parameters to check the scaling as a function of temperature and voltage. Therefore, we fit the Hikami-Larkin-Nagaoka formula \cite{hikami1980spin}
\begin{equation}\label{eq:hln}
    \sigma(B) - \sigma(0) = -\frac{\alpha e^2}{2\pi^2\hbar}\left[\ln \frac{B_\phi}{B} - \psi\left(\frac{1}{2} + \frac{B_\phi}{B} \right) \right].
\end{equation}
Here, $\alpha$ is a prefactor related to the number of parallel channels contributing to the quantum correction \cite{garate2012weak}, and $B_\phi = \hbar/4eL_\phi^2$. We use both $\alpha$ and $L_\phi$ as tracking parameters. Although the HLN equation does not capture the full physics of EEI and WAL at finite temperatures \cite{lu2014finite}, it provides a good fit to the data and thereby allows for comparing traces taken at different $V,\ T$, without studying in detail what the magnitude of both parameters implies.

From the HLN fit (performed on the data up to 1 T), we find that $L_\phi$ and $\alpha$ are suppressed both by $T$ and $V$. We plot $L_\phi$, $\alpha$ versus $\epsilon = eV$ in Fig.~\ref{fig:VT_curves}(b). To compare whether the quadratic relation between $V$ and $T$ holds for the magnetoresistance as well, we overlay $L_\phi$, $\alpha$ as function of $T^2$ and find that the scaling approximately matches the voltage-dependent dataset.

Having found a conversion between temperature and voltage in magnetotransport data, we next solidify whether the same temperature-voltage correspondence holds for the EEI and WAL contributions identified in differential conductance measurements. To do so, we search for points in temperature and voltage where the conductivity corrections are equal.
The process is illustrated in Fig. S7 \cite{TV_app}. Firstly, we find points where the conductivity change with respect to the maximum conductivity ($\Delta \sigma$) is equal in temperature ($I_\mathrm{DC} = 0$ $\mu$A) and voltage ($T = 2.5$ K) dependent data. We plot the resulting $T(V)$ curve in Fig.~\ref{fig:VT_curves}(c) at both 0 T (EEI and WAL) and 9 T (EEI only).
Secondly, we study the difference between differential conductances measured at 0 T and 9 T, which removes the temperature-dependent background and EEI contribution, leaving the WAL. This $T(V)$ curve is shown in Fig. \ref{fig:VT_curves}(c) as well. The error bars, $\delta T(V)$, are obtained from the (fixed) noise level in $\Delta \sigma$, $\delta(\Delta \sigma)$. The error in the obtained temperature at each voltage is then dependent on the slope of $\Delta \sigma$ as function of $T$ at the equivalent temperature, and is given by $\delta T(V) = \delta(\Delta\sigma)/(\partial \Delta \sigma/\partial T)$.

Both datasets follow $V = AT^2$, where $A \approx 2.24\cdot10^{-3}$ V/K$^2$. The quadratic relation is in line with the linear FWHM \cite{zbrp_app} and was also found in previous research \cite{nandi2018logarithmic}. 
Based on this observation, we can relate all contributions in the differential resistance to the same temperature-voltage correspondence, and we conclude that both EEI and WAL are suppressed via Joule heating upon applying a bias voltage.

\section{Dephasing mechanism}
In the magnetoresistance analysis, we have employed the HLN formula. However, in order to fully incorporate finite-temperature effects of weak antilocalization and electron-electron interaction, a more involved model containing both EEI and quantum interference (WAL or WL) has been created by Lu and Shen (LS) \cite{lu2014finite}. As a function of magnetic field, changes in the quantum interference correction overshadow changes in the EEI correction, whereas in the temperature dependence both corrections contribute. Although the LS model does not explicitly include bias voltage as a parameter, it does include it implicitly, because the temperature changes due to Joule heating at finite $V$. This way, we can model the differential conductance as a function of $V$ in the LS model by substituting the temperature-voltage correspondence $V = 2.24\cdot 10^{-3}T^2$, obtained in Fig.~\ref{fig:VT_curves}(c).

The governing parameter of the LS model is $\Delta/E_f$, where $\Delta$ is a finite gap opened by a mass term. Surface state transport corresponds to $\Delta/E_f \rightarrow 0$, resulting in weak antilocalization, and a Fermi level close to the band bottom in a gapped structure corresponds to $\Delta/E_f \rightarrow 0$. 
Using the measured {charge-carrier density} $n_s \sim 10^{13}$ cm$^{-2}$ and $v_F \sim 4.7 \cdot 10^5$ m/s \cite{mulder2022spectroscopic}, in case of a linear dispersion we estimate $E_F \sim 245$ meV. This value is the same order of magnitude as the estimated bulk band gap in (Bi$_{1-x}$Sb$_x$)$_2$Te$_3$ \cite{zhang2011band}, and therefore transport is likely a combination of surface and bulk states, requiring multiple channels in the LS model.

Even when including surface and bulk contributions, the LS model will result in a single value for the slope $\dd \sigma/\dd(\ln T)$. To explain our measured differential conductance, this slope should change from positive to negative values as function of $T, V$. The slope in the LS model hinges on a key assumption: $L_\phi \sim T^{-p/2}$ for topological insulators, as experimentally verified \cite{peng2010aharonov, checkelsky2011bulk}. However, $L_\phi$ extracted from our data does not follow this power law, see Fig.~\ref{fig:VT_curves}(b) {and Fig.~S8 \cite{LS_app}}. Deviations from $L_\phi \sim T^{-p/2}$ have been observed before. For instance, the bulk of a topological insulator can dephase due to variable range hopping whereas Nyquist dephasing dominates edge state transport \cite{banerjee2018spatially} or charge puddles in the bulk could cause dephasing \cite{liao2017enhanced}. If $L_\phi$ does not scale $\sim T^{-p/2}$, the LS model cannot be applied straightforwardly \cite{tkavc2019influence}.

If we use our fitted $L_\phi$ as model input and substitute $V = 2.24\cdot 10^{-3}T^2$, we can obtain qualitatively similar differential conductance curves to the data \cite{LS_app}. The slower decay of phase coherence length with temperature allows the decay in EEI contributions to dominate at low temperatures, after which the slope $\dd \sigma/\dd(\ln T)$ crosses over from positive to negative. 

Before drawing a conclusion on whether the LS model exactly describes our data with altered phase coherence length, more research into the dephasing mechanisms in our (Bi$_{1-x}$Sb$_x$)$_2$Te$_3$ films is required, for instance by studying quantum oscillations in devices of reduced size \cite{peng2010aharonov, checkelsky2011bulk}. Close attention should be paid to whether the LS theory would still be valid in the observed transport regime.

Apart from the dephasing mechanisms, the difference between lattice and electron temperature as a result of Joule heating can be discussed as well. 
The question remains: would such a high ($\sim$ 20 K) difference between electron and lattice temperature be attainable? Fully answering this question requires solving the theory by Abrahams and Anderson using the heat capacity of BST films \cite{anderson1979possible}. The exponent $2/(2+p)$ is derived by considering the temperature dependence of the free electron heat capacity, which differs from topological insulator surface states. Therefore we cannot attribute a value of $p$, even though we found $V \sim T^2$. Nonetheless, the low electron-phonon coupling in topological surface states \cite{pan2012measurement} makes an elevated electron temperature likely.

\subsection{Lattice temperature}
We have estimated the electron temperature as function of bias voltage based on resistance measurements. To verify this estimate, alternative temperature measurement techniques are desirable. While noise thermometry, used for probing electron temperatures \cite{muller2021electron}, is not available to us, we can still examine the lattice temperature. This way, we can probe whether only the electron temperature increases or if heat transfer to the lattice is significant.
\begin{figure*}
    \centering
    \includegraphics{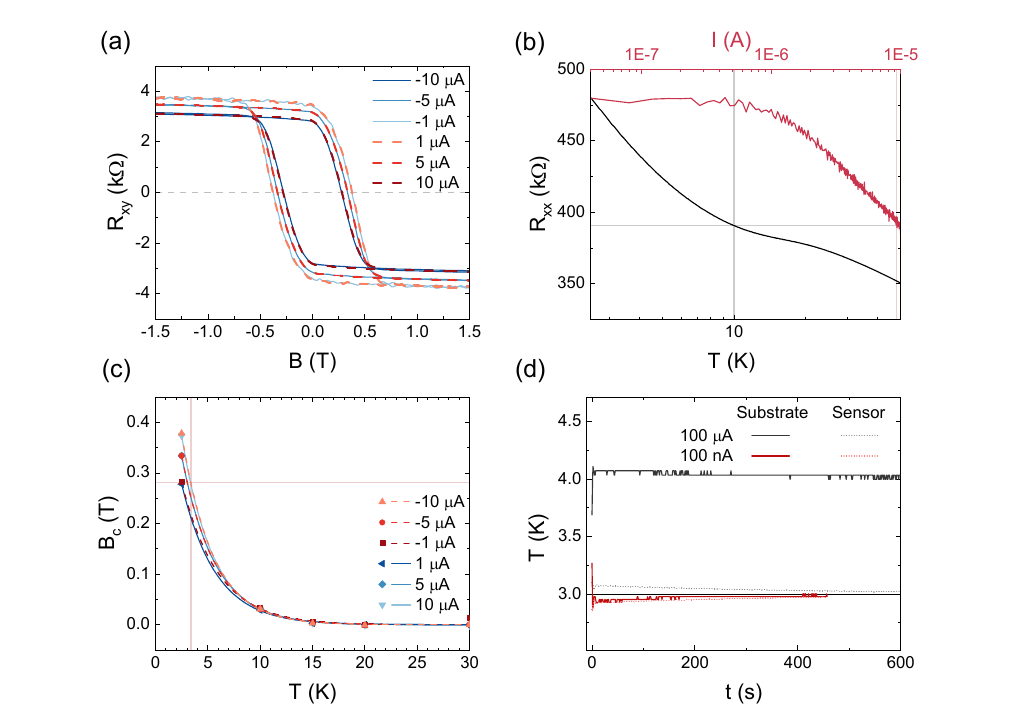}
    \caption{(a) $R_\mathrm{xy}$ in a magnetically doped V$_y$(Bi$_{1-x}$Sb$_x$)$_{2-y}$Te$_3$ Hall bar (D3) at $T = 2.5$ K. The data is hysteretic: the intrinsic magnetization changes sign when the external magnetic field exceeds the coercive value $B_c$. $B_c$ is suppressed when increasing the DC bias current. (b) $R_{xx}$ as function of $T$ measured at $I_\mathrm{DC} = 1$ $\mu$A (black) and $\dd V_{xx}/\dd I$ as function of $I_\mathrm{DC}$ at $T = 2.5$ K (red). At $I_\mathrm{DC} = 10$ $\mu$A the equivalent electron temperature is 10 K. (c) $B_c$ versus temperature for a range of $I_\mathrm{DC}$. The lattice temperature increases up to 3 K for high bias currents when comparing the $\pm 10$ $\mu$A to the $\pm 1$ $\mu$A dataset. (d) Substrate temperature change due to Joule heating in a (Bi$_{1-x}$Sb$_x$)$_{2-y}$Te$_3$ Hall bar device (D4) at T = 3 K, and corresponding temperature measured by PPMS puck thermometer. The substrate temperature is calculated from the resistance change in a neighboring Hall bar device, which is measured using standard low-bias lock-in techniques ($I = 100$ nA).}
    \label{fig:5}
\end{figure*}

To study the lattice temperature while applying a bias, we turn to vanadium-doped BST (VBST) films. The crystal structure is equal to BST, where vanadium doping resides on either a bismuth or antimony site. {In previous research, similar non-Ohmic contributions to our observations have been found in VBST \cite{nandi2018logarithmic}, where an equivalent scaling of $V \sim T^2$ was found as well. Because of the similarity between the observed mechanisms, we use magnetically-doped VBST to further investigate the lattice- and electron temperature.}. 

Our VBST is deposited on sapphire substrates, likewise to BST, {with vanadium atoms residing on the Bi/Sb lattice sites. Because of the similar crystal structure, {charge-carrier densities} ($n_\mathrm{s,\ VBST} = 1\cdot10^{13}$ cm$^{-2}$), and previous research on VBST \cite{nandi2018logarithmic}, we assume that we can draw comparable conclusions on heating when comparing VBST to BST.} Contrary to non-magnetic BST, VBST is ferromagnetic below the Curie temperature. In VBST Hall bars, $R_{xy}$ is hysteretic due to the intrinsic magnetization changing sign when the external field exceeds the coercive value $B_c$ in Fig.~\ref{fig:5}(a). In magnetic topological insulators, the magnetism can be either mediated by band electrons (Van Vleck) or by itinerant carriers (Ruderman–Kittel–Kasuya–Yosida, or RKKY) \cite{wang2023evolution}.  If the magnetism is of the Van Vleck type, as has been shown in previous transport experiments on our VBST films \cite{wielens2021axion}, we expect the coercive field to be independent of the electron temperature. 

In Fig.~\ref{fig:5}(b) we show $R_{xx}$ as a function of temperature (using $I = 1\ \mu$A) and differential resistance as function of bias current (at $T = 2.5$ K). From this figure, the temperature should increase to 10 K at the maximum bias. Meanwhile, in Fig.~\ref{fig:5}(c) we characterize $B_c$ as a function of temperature, for both low and high bias currents. At 2.5 K, the lattice temperature changes to 3 K at most, based on comparing the low-temperature $B_c$ at $\pm10$ $\mu$A to the temperature-dependent $B_c$ at $\pm1$ $\mu$A (interpolated by fitting an exponential curve). {Figures~\ref{fig:5}(b) and (c)} emphasize that Joule heating indeed causes a larger temperature difference in electron temperature than in lattice temperature.
We remark that the VBST films will have a larger discrepancy between temperature-dependent data at low bias and voltage-dependent data at low temperature, because the intrinsic magnetization has a large effect on transport and is not heavily influenced by Joule heating.

Another method of probing the lattice temperature, applicable to non-magnetic devices as well, is shown in Fig.~\ref{fig:5}(d). Here we track the resistances of a main device (D4 in Table~\ref{tab:devices}) and a neighboring Hall bar device on the same sapphire substrate. The inter-device spacing is 1.25 mm. Sapphire has a high thermal conductivity \cite{crystek}, so assuming negligible thermal gradients between Hall bars and substrate we can approximate the lattice temperature to be comparable in both devices. Using a low excitation current we track the resistance of the neighboring device while applying a $I_\mathrm{DC} = 100$ $\mu$A bias on the main device ($R = 28$ k$\Omega$, so the heating power is similar to measurements on D1). Using this method we observe an increase of 1 K, in contrast with the 0.1 K increase read out by the RuOx temperature sensor mounted on the PPMS measuring probe. Although this is not an exact method of measuring the lattice temperature, we can estimate that the lattice temperature in the measurements on D1 and D2 will not have changed significantly upon applying a DC bias.

\subsection{Alternative dephasing mechanisms}
Alternative mechanisms suppressing weak antilocalization have been proposed in literature as well. Firstly, suppression of weak antilocalization has been linked to dynamic nuclear polarization \cite{jiang2020dynamic}. The distinct difference with our dataset is that we have not found a time dependence linked to WAL broadening: the measured magnetoresistance was independent of the measuring history (within experimental resolution). 

Secondly, the effect of a DC bias on the quantum corrections in graphene has been explained as additional `dispersion decoherence' in the material, not requiring Joule heating \cite{somphonsane2020universal}. The dispersion decoherence terms become apparent when deriving the WL/WAL correction beyond linear response theory (which cannot be applied to the nonequilibrium regime). The conductance curves collapse to a single curve as well when plotted as a function of $eV/k_\mathrm{B} T$, similar to our results. However, in graphene, the applied bias voltage to suppress the zero bias resistance features was too small to reach the required electron temperature, so Joule heating was excluded.

\section{Conclusion}
We have shown that the non-Ohmic components in the differential resistance can be attributed to EEI and WAL being suppressed by Joule heating. Although these contributions both scale differently as a function of temperature (or voltage), the temperature-voltage correspondence is equal for both. The temperature- and voltage-dependent transport properties do not map one-to-one, because of a lattice temperature-dependent background. 

The results presented here form a new perspective on non-Ohmic signatures in topological insulators. Zero-bias resistance peaks have been observed numerous times, for instance when superconductivity is suppressed in three-dimensional topological insulator Josephson junctions \cite{stehno2017conduction, janssen2024characterization}. We conclude that these signatures can be caused by an elevated electron temperature at finite bias, which suppresses quantum corrections to the conductivity. This observation is compatible with existing theory of quantum corrections at finite temperatures in the LS model.
Therefore, it is crucial to consider Joule heating when studying non-Ohmic signatures in topological insulators at low temperatures, as it plays a significant role in influencing these transport properties.

\section{Acknowledgements}
This research was supported by a Lockheed Martin Corporation Research Grant.\nocite{data}
\vfill

\FloatBarrier
\bibliography{references}

\FloatBarrier
\widetext
\clearpage
\supplementarysection
\subsection{{Device geometry}}
\begin{figure}[h]
    \centering
    \includegraphics{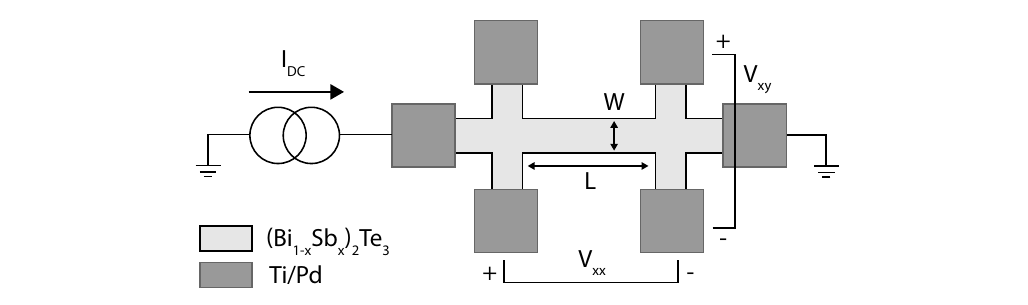}
    \caption{{Schematic depiction of the device layout and measurement configuration.}}
    \label{fig:device_schem}
\end{figure}
\subsection{{Hall data}}\label{sec:testrefapp}
\FloatBarrier

\begin{figure}[h]
    \centering
    \includegraphics{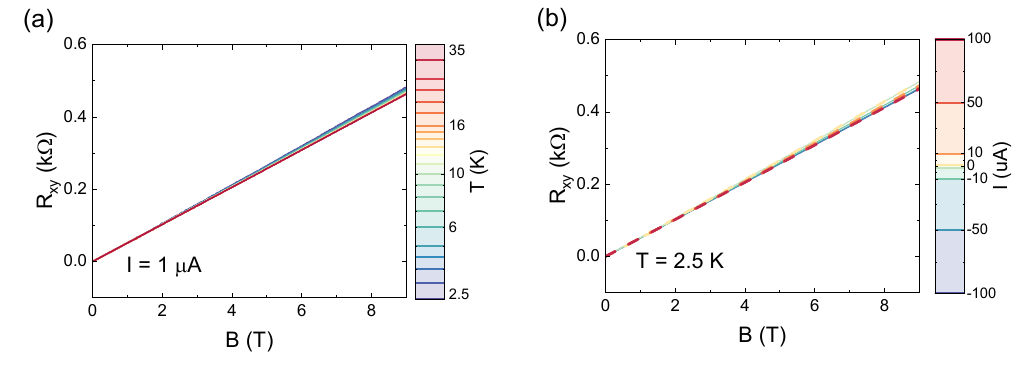}
    \caption{$R_{xy}$ as function of magnetic field for a range of (a) temperatures (at $I_\mathrm{DC} = 1$ $\mu$A) and (b) voltages (at $T = 2.5$ K). {The traces obtained using $I_\mathrm{DC}>0$ and $I_\mathrm{DC}<0$ in (b) overlap; results for $I_\mathrm{DC}>0$ in (b) are plotted using dashed lines for clarity.} The data has been antisymmetrized in magnetic field.}
    \label{fig:Rxy}
\end{figure}
\FloatBarrier
{In the main text, we studied the non-Ohmic signatures in longitudinal resistance ($R_{xx}$). In Fig. \ref{fig:Rxy}, we provide the Hall data ($R_{xy}$) corresponding to the magnetotransport data shown in Fig. \ref{fig:MR_IvsT} of the main text. The data is anti-symmetrized in magnetic field.}

 Another method to distinghuish localization and interaction effects is by studying the Hall resistance. The Hall \textit{conductivity} is unaffected by electron electron interaction, so upon inverting the conductivity tensor one finds \cite{lee1985disordered, altshuler1980magnetoresistance, kuntsevich2016low}
\begin{equation}\label{eq:Hallcorr}
    \frac{\Delta \rho_{xy}}{\rho_{xy}}  = 2\frac{\Delta \rho_{xx}}{\rho_{xx}}
\end{equation}

\begin{figure}
    \centering
    \includegraphics[width=0.6\textwidth]{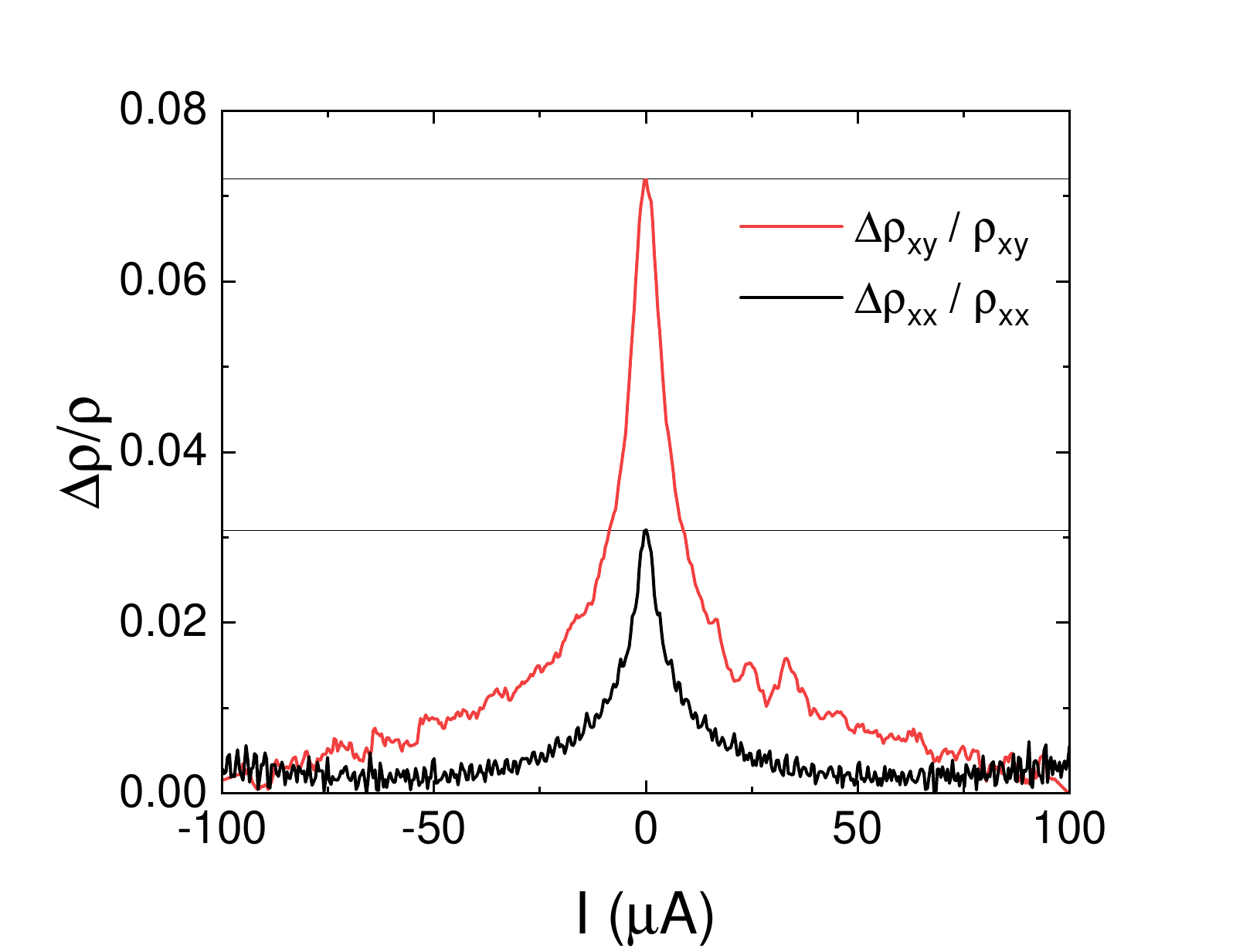}
    \caption{Relative magnitude of the zero-bias resistance peak at $B = 9$ {T and $T = 3$ K. Horizontal lines provide guides to the eye}. The approximate factor 2 difference is in line with Eq. (\ref{eq:Hallcorr}). Data from device D1.}
    \label{fig:delta_rho}
\end{figure}
\FloatBarrier
\clearpage
\subsection{Filtering effects}
\begin{figure}[h!]
    \centering
    \includegraphics[width=\textwidth]{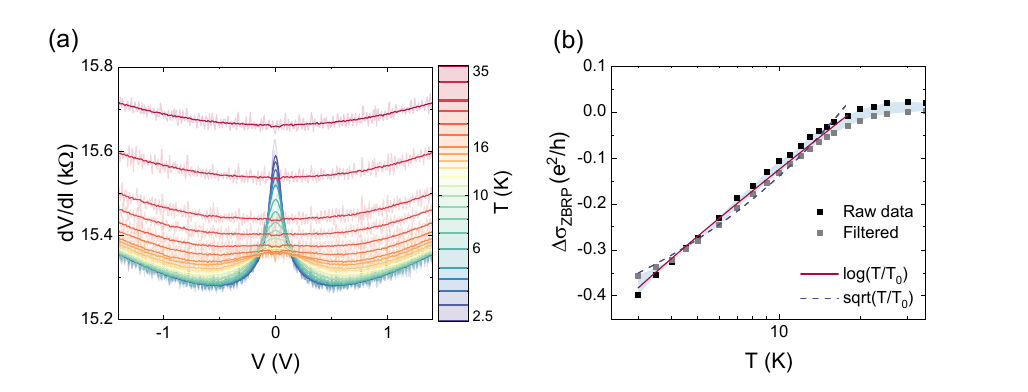}
    \caption{(a) Filtered (solid) and unfiltered (transparent) differential resistance for a range of temperatures at $B = 0$ T. At low bias, the filtering suppresses the differential resistance. (b) Conductivity correction corresponding to the zero-bias resistance peak for filtered and unfiltered data. The logarithmic dependence is apparent in the raw data, the suppression of the ZBRP due to filtering distorts the dataset. Data from device D1.}
    \label{fig:filters}
\end{figure}
\begin{figure}[h!]
    \centering
    \includegraphics[width=0.5\textwidth]{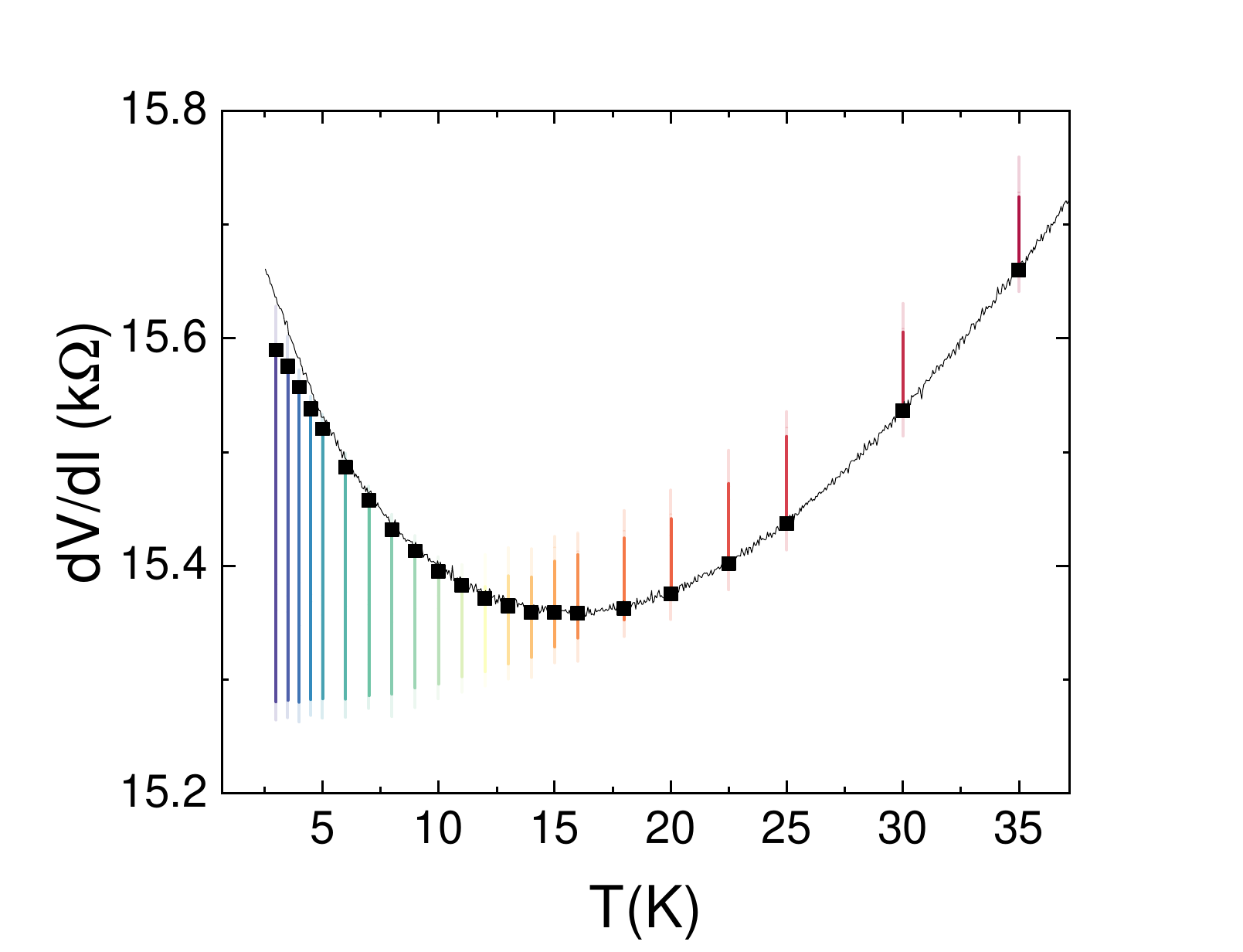}
    \caption{Data from Fig. \ref{fig:filters}(a) as function of temperature (colored lines). The dataset is accompanied by the zero-bias differential resistance obtained from the filtered dataset (data points) and resistance measured while cooling down the sample ($R(T)$) at $I_\mathrm{DC} = 1$ $\mu$A (solid black line). At low temperatures, the unfiltered (transparent) lines give the most accurate results when comparing to $R(T)$, whereas at high temperatures the filtered (solid) dataset is more accurate. Data from device D1.}
    \label{fig:cooldown}
\end{figure}
\FloatBarrier
\clearpage
\subsection{Universal scaling of the zero-bias resistance peak}\label{sec:fwhm_theory}
{In the main text, we extract the full-width at half-maximum (FWHM) of the zero-bias resistance peak. Figure \ref{fig:sigma_ZBRP_2}(c) shows that the FWHM has universal scaling ($V_\mathrm{FWHM}\propto T$). Here, we show how a logarithmic scaling of the resistance with $V$ and $T$, combined with universal scaling of the $V_\mathrm{FWHM}$, implies that a correspondence of $V \propto T^2$ holds.}

\begin{figure}[h!]
    \centering
    \includegraphics[width=\textwidth]{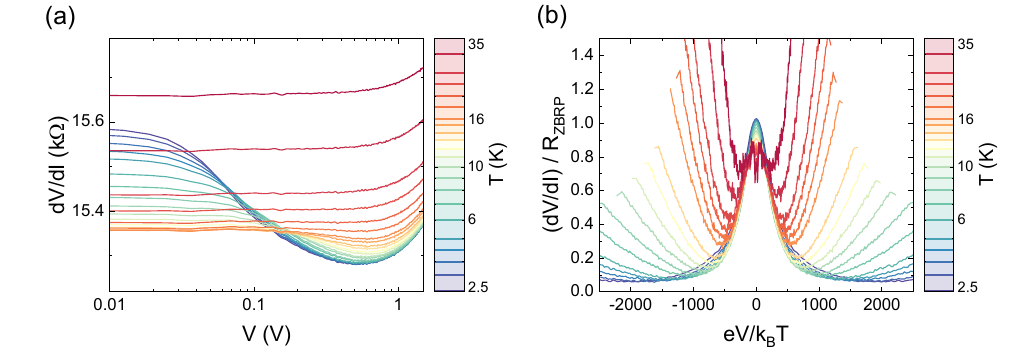}
    \caption{(a) Data from Fig.~\ref{fig:sigma_ZBRP_1} (D1) on logarithmic voltage axis, showing a logarithmic decay of the ZBRP and subsequent increase of the resistance. (c) Data from  Fig.~\ref{fig:sigma_ZBRP_1} (D1) normalized to ZBRP magnitude as a function of $eV/k_\mathrm{B}T$. The data around zero bias collapse to a single curve. (d) Full width at half maximum extracted from Fig.~\ref{fig:sigma_ZBRP_1}, with a linear fit.}
    \label{fig:suppl_zbrp}
\end{figure}

Starting from a value $R_0$ at low temperature and voltage, we calculate the scaling when changing either variable. As function of temperature, the resistance scales as
\begin{equation}\label{eq:RT}
    {R(T, V=0) = R(T_0, V=0)} - A \ln\frac{T}{T_0}\,
\end{equation}
{where $R(T_0, V = 0)$ equals the resistance at a reference temperature $T_0$ and $A$ is a proportionality factor. The differential resistance, see Fig.\ref{fig:suppl_zbrp}(a), saturates at low voltage and decreases logarithmically beyond a reference value $V_0$. Hence we can write, for $V > V_0$,}
\begin{equation}\label{eq:RV}
    {R(T, V) = R(T, 0)} - B \ln\frac{V}{V_0},
\end{equation}
{where $B$ is a proportionality factor. The full width at half maximum corresponds to the voltage where $R$ equals half its maximum, the low-voltage value at finite temperature, or using Eq. \ref{eq:RT}:}
\begin{align*}
R(T, V_\mathrm{FWHM}) = \frac{R(T_0, 0) - A\ln{\frac{T}{T_0}}}{2}.
\end{align*}
{$R_\mathrm{FWHM}$ is measured at finite temperature and voltage. We now use Eqs. (\ref{eq:RT})-(\ref{eq:RV}) and find} \
\begin{align*}
R(T, V_\mathrm{FWHM}) = R(T_0, 0) - A\ln{\frac{T}{T_0}} - B\ln{\frac{V_\mathrm{FWHM}}{V_0}} = \frac{R(T_0, 0) - A\ln{\frac{T}{T_0}}}{2}
\end{align*}
and hence
\begin{equation}\label{eq:fwhm}
    \frac{V_\mathrm{FWHM}}{V_0} = \left(\frac{T}{T_0}\right)^{A/2B}.
\end{equation}
We found that the {FWHM} scales similarly with voltage and temperature if $V \propto T^2$, equivalent to $A = 2B$. Equation \ref{eq:fwhm} then reduces to the universal scaling ($V_\mathrm{FWHM} \propto T$) observed in Fig. \ref{fig:sigma_ZBRP_2}(c) of the main text.

\FloatBarrier
\clearpage
\subsection{Temperature-voltage correspondence}
\begin{figure}[h!]
    \centering
    \includegraphics[width=0.5\columnwidth]{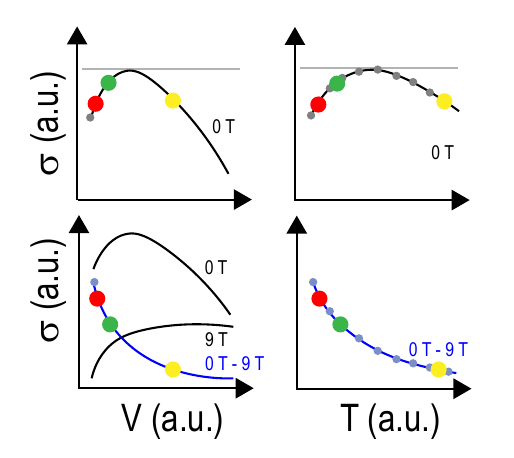}
    \caption{Method of finding the temperature-voltage correspondence. At 0 T, both EEI and WAL contribute. To reduce the effects of the temperature-only dependent background, we consider the relative change $\sigma(V) - \sigma_\mathrm{max}$. The zero-voltage value from (a) is extracted for each temperature in (b), after which the comparison between temperature and voltage effects is made. The $T(V)$ curve consists of the $T,V$ values where the curves in (a) and (b) are equal (corresponding to red/green/yellow dots). A similar analysis is performed to exclude EEI (c-d), where the difference between 0 T and 9 T data is considered (see also Fig. \ref{fig:3_WAL_sigBG_1}).}
    \label{fig:methods}
\end{figure}

\FloatBarrier
\clearpage
\subsection{Lu Shen model for alternate dephasing mechanisms}

\begin{figure}[h!]
    \centering
    \includegraphics[width=0.5\textwidth]{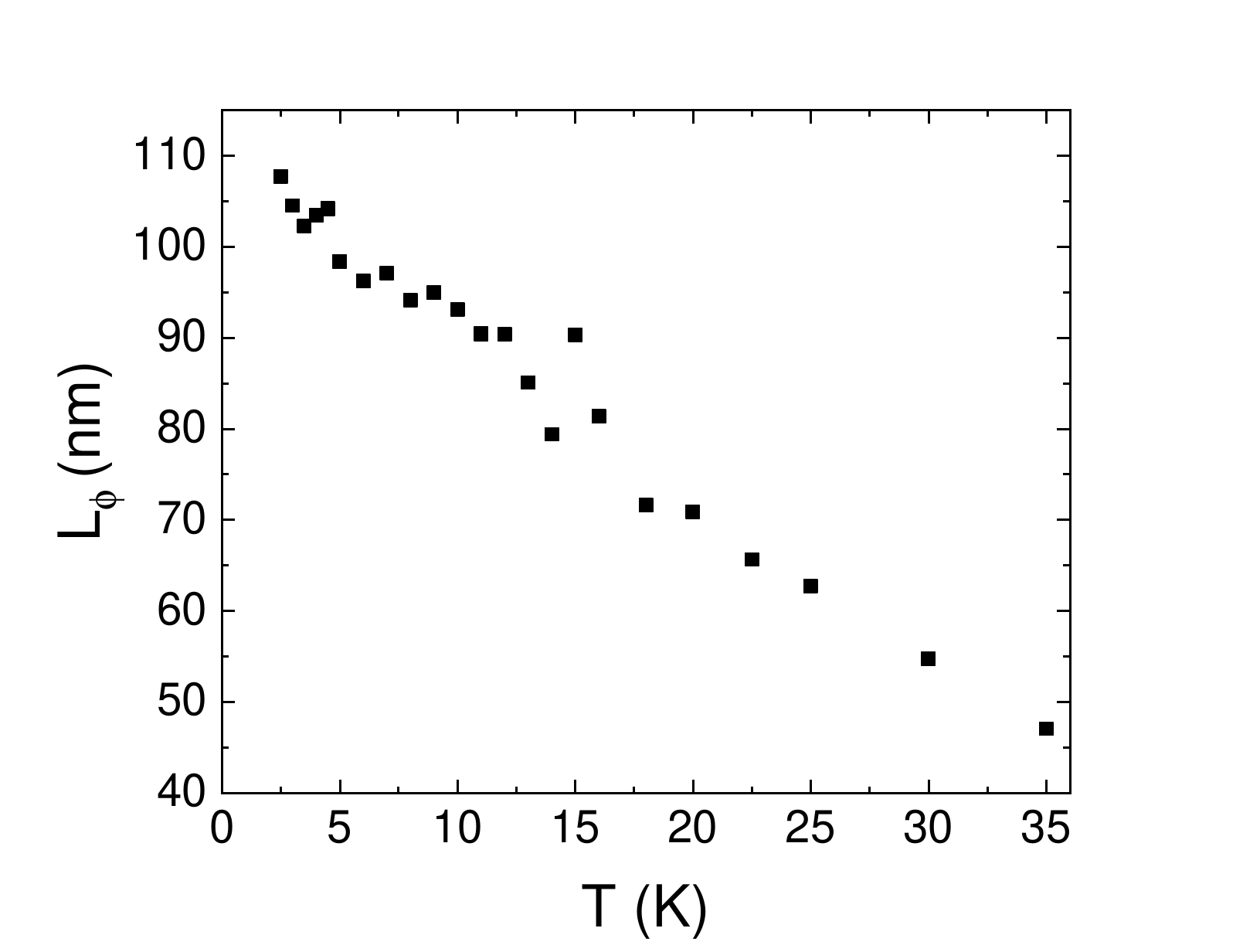}
    \caption{{Phase coherence length $L_\phi$ obtained from the HLN fit at varying temperatures, using the magnetoresistance data in Figs. \ref{fig:MR_IvsT}(a). The results do not follow the power law dependence ($L_\phi \sim T^{-p/2}$) assumed in the Lu Shen model \cite{lu2014finite} and observed in other topological insulators \cite{peng2010aharonov, checkelsky2011bulk}, and hence the LS model does not apply directly \cite{tkavc2019influence}. The phase coherence length in a topological insulator might deviate from a power-law dependence because dephasing mechanisms can vary. The bulk of a topological insulator may experience dephasing through variable-range hopping, while edge state transport is primarily affected by Nyquist dephasing \cite{banerjee2018spatially}. Additionally, dephasing in the bulk can arise from charge puddles \cite{liao2017enhanced}. If the phase coherence length $L_\phi$ does not follow a scaling of $\sim T^{-p/2}$, the LS model cannot be directly applied \cite{tkavc2019influence}.    
    Data from device D1, see also Fig.~\ref{fig:VT_curves}(b).}}
    \label{fig:suppl_Lphi}
\end{figure}

\begin{figure}[h]
    \centering
    \includegraphics{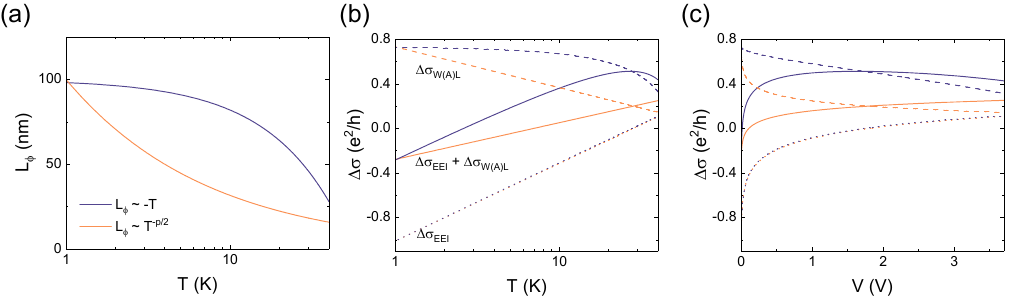}
    \caption{Lu Shen model using $\varepsilon_r = 100$, $\gamma = 3$ eV \r{A}, $\theta = 0.5 \pi$ and $B = 0$ T \cite{lu2014finite}. (a) $L_\phi$ in a model similar to our dataset ($\propto -T$) and as used in model by Lu and Shen ($\propto T^{-p/2}$, $p = 1$). (b) $\Delta \sigma$ as function of $T$ when using the different dephasing models from (a). Dotted lines show that the influence of $L_\phi$ on $\Delta \sigma_\mathrm{EEI}$ is negligible, whereas dashed lines show that $L_\phi$ is a dominant factor in the temperature dependence of $\Delta \sigma_\mathrm{WAL}$. In the total $\Delta \sigma = \Delta \sigma_\mathrm{EEI} + \Delta \sigma_\mathrm{WL/WAL}$ a nonmonotonous slope is visible when the experimental $L_\phi$ is used. (c) Bias voltage dependence of $\Delta \sigma$ when using $V = 2.24\cdot 10^{-3}T^2$.}
    \label{fig:lushen}
\end{figure}
\end{document}